# Evaluation of algorithms for correction of transcranial magnetic stimulation induced artifacts in electroencephalograms

Panteleimon Vafeidis[a,*], Vasilios K. Kimiskidis[b], Dimitris Kugiumtzis[c]

[a]*Bernstein Center for Computational Neuroscience Berlin, Berlin, Germany*

[b]*Laboratory of Clinical Neurophysiology, Medical School, Aristotle University of Thessaloniki, Thessaloniki, Greece*

[c]*Department of Electrical and Computer Engineering, Faculty of Engineering, Aristotle University of Thessaloniki, Thessaloniki, Greece*

*E-mail: vafeidis@campus.tu-berlin.de

## Abstract

Transcranial magnetic stimulation combined with electroencephalography (TMS-EEG) is widely used to study the reactivity and connectivity of brain regions for clinical or research purposes. The electromagnetic pulse of the TMS device generates at the instant of administration an artifact of large amplitude and a duration up to tens of milliseconds that overlaps with brain activity. Methods for TMS artifact correction have been developed to remove the artifact and recover the underlying, immediate response of the cerebral cortex to the magnetic stimulus. In this study, three such algorithms are evaluated. Since there is no ground truth for the masked brain activity, pilot data formed from the superposition of the isolated TMS artifact on the EEG brain activity are used to evaluate the performance of the algorithms. Different scenarios of TMS-EEG experiments are considered for the evaluation: TMS at resting state, TMS inducing epileptiform discharges and TMS administered during epileptiform discharges. We show that a proposed gap filling method is able to reproduce qualitative characteristics and in many cases closely resemble the hidden EEG signal. Finally, shortcomings of the TMS correction algorithms as well as the pilot data approach are discussed.

## Introduction

Transcranial magnetic stimulation (TMS) is a noninvasive method for brain stimulation with electromagnetic pulses delivered via stimulation coils [1]. The pulse has short duration (< 1 ms) and peak amplitude up to 3 Tesla. Short duration-high frequency induced electric currents are able to penetrate the cell membranes and depolarize pyramidal cells and interneurons. Depending on the equipment, the depth of the stimulation can be up to 6 cm [2,3]. The reaction of the cortex can be recorded by combining TMS and electroencephalography (TMS-EEG) [4] and depends on different factors such as: the precise site of stimulation [5], the direction of the coil, the type of the coil [6], the power of the pulse [7,8], the form of the pulse [9], the state of the cortex [10,11], and the level of consciousness [12]. The reaction occurs first at the stimulation area and then spreads to more distant areas as the activation is transmitted along existing neural pathways. Thus,

TMS-EEG is an excellent tool to measure connectivity between distinct brain areas and reactivity to a specific stimulation pattern [13]. TMS-EEG has been utilized for diagnostic [14] and therapeutic purposes [15]. Repetitive TMS (rTMS) can act as a modulator of neuronal activation and leads to reorganization of cortex connections with numerous therapeutic implications [4].

A key issue in the preprocessing of TMS-EEG recordings is the correction of artifacts generated by the electromagnetic pulse of the TMS device and the unintended activation of physiological systems other than the brain *per se* (muscle and trigeminal nerve activation, eye movements) [4,6]. The artifact is observed in the EEG as fluctuations of several orders of magnitude larger than electrophysiological activity. Though several techniques have been suggested and are in use for reducing this effect, e.g. a sample-and-hold circuit [16,17,18], improved EEG amplifiers such as direct current coupled amplifiers [19] and use of TMS-compatible EEG electrodes [20,21], the TMS-related artifact remains in the EEG recording, is relatively large in amplitude and has a duration up to 20 ms. Thus, early TMS evoked potentials (i.e. at a latency ≤ 20ms) are masked by the TMS-related artifact. Complete removal of channels heavily contaminated by the artifact from the analysis would mean that valuable information would be lost (since these channels would be exactly the ones close to the stimulation site), e.g. this could lead to a systematic error in source estimation [2].

The analysis of TMS-EEG requires the offline correction of the TMS related artifact, termed simply TMS artifact. Blind source separation methods have been proposed to separate cortical and non-cortical sources and thus remove the TMS artifacts, such as techniques based on independent component analysis (ICA) [3,4,22,23] and principal component analysis (PCA) [24,25], and the so-called signal-space projection [26,27]. ICA-based TMS artifact correction algorithms have been built in the FieldTrip toolbox [28], as well as the two recent Matlab modules for TMS-EEG processing, TMSEEG [29] and TESA [30]. In a different approach, rather than attempting to identify and remove the artifact from the signal, the whole signal affected by the artifact is treated as corrupted and considered as a gap to be filled by an interpolation method, such as linear or cubic interpolation [31,32]. A more advanced interpolation approach is proposed in [33], filling the gap from the weighted forward-backward prediction of a local state space model.

The immediate cortex activity evoked by the TMS is not known as it is masked by the TMS artifact, and thus there is no ground truth to be used as reference for the evaluation of TMS artifact correction algorithms. For the evaluation of the algorithms, qualitative characteristics of the corrected signal are used, e.g. based on its time-frequency representation [34].

In this study, we evaluate three of the artifact correction algorithms discussed above. The first two identify and remove the artifact from the signal, i.e. the FastICA algorithm [35] and the separation making use of PCA and the source to sensor forward model, termed here PCA model [2], and the advanced gap filling method in [33]. For the evaluation, we first estimate the artifact from TMS-EEG recordings at rest, obtained by averaging over several iterations of stimulation. The artifact is then added to three types of pilot high-density EEG signals: resting EEG, at the onset of an epileptiform discharge (onset ED) and in the course of an epileptiform discharge (mid-ED). The correction algorithms are applied to the resulting signals. The signal derived from the superposition of TMS artifact and the initial EEG signal lacks the possible interplay of artifact and underlying neuronal activity but allows for using the initial EEG as reference for evaluating the corrective capacity of each algorithm.

# Methods

The structure of the study is as follows. First, the three TMS artifact correction methods are discussed, and then the procedure for generating EEG signals added with TMS artifact as well as the estimation and evaluation of the correction methods are presented.

## TMS-related artifact correction methods

### FastICA

FastICA is a time effective algorithm implementing independent component analysis (ICA). Possible pre-processing steps are data centering, whitening and dimensionality reduction, the latter being done by PCA in FastICA. The rationale of ICA is to separate independent sources being mixed in the observed multichannel EEG signal [36]. For this, the statistical independence is maximized to find the independent components (ICs) using a procedure that maximizes the non-Gaussianity. FastICA uses the so-called fast fixed-point algorithm to run faster and projection pursuit to find the ICs. The algorithm can be configured to produce the ICs one at a time (deflation approach) or converge to the set of ICs (symmetric approach). Statistical independence is a very strong condition in ICA and requires an infinite length of data to be verified. An acceptable lower limit for the length of a signal of *N* channels is $3N^2$. For shorter EEG signals, PCA is used to reduce the dimension *N*.

ICA generally outputs the ICs in random order. As FastICA uses PCA for pre-processing, ICs are given in decreasing variance (power) order and since the artifact has much larger variability than the physiological signal, the ICs corresponding to the artifact come first. Specifically, recognition can be done in two ways. Firstly, by observing the ICs themselves, and looking for large fluctuations at the time of the artifact that decay to zero when there is no artifact. These ICs can be identified automatically, by centering each IC around its average value, taking the absolute value and deciding for rejection according to a given threshold. The reconstructed artifact-free signal is then obtained by the remaining ICs. The second way is more complex and is based on the observation of the topography matrix that represents the spatial distribution of each IC. Knowing the stimulation site, the distribution of the artifact on the scalp can be estimated, identified and finally removed. ICA and FastICA implementations are standard in EEG processing software packages, such as EEGLAB [37] and Fieldtrip [28].

### PCA model

The method introduced in [2] attempts to separate the neuronal activity signal from artifact activity as ICA, but unlike the latter it makes no assumption about spatial or temporal independence of brain and artifact activities. However, it relies on the construction of a source model consisting of brain and artifact topographies. The source model and a linear inverse operator decompose the data into a linear combination of artifact and neural signals and finally the artifact signals are subtracted from the data. The model for the brain topographies requires that the head model and the electrode positions are known. Then activating a dipole with a known position produces a specific topography of activity at the electrode positions on the head surface. The information of this linear and fixed-to-time activation relationship is forwardly produced and stored in the lead field matrix. The artifact topographies are obtained from PCA decomposition of the averaged artifact using the minimum possible sources. For TMS–EEG data, this is the early signal (< 20 ms) that primarily represents the TMS artifact.

If the topography of the neuronal sources is known in advance, models of previous studies can be used. In the general case, however, this is not feasible. So, this algorithm uses a two-step iterative process to approximately model the neuronal sources. In the first step, neuronal activity is represented by a surrogate model, with 15 sources distributed around all major cortical regions. This model is capable of approximately reproducing any recorded activity on the surface, with an appropriate activation of each source. The artifact is corrected using the model derived by PCA and the surrogate neuronal model, and the corrected signal is used to build a new source model specifically for neuronal activity. In the second step, the correction is done using this improved model of neuronal sources in synergy with the lead field matrix for the artifact sources already calculated. So, the positions of the sources are not forced to be those of the surrogate model.

The quality of separation of neuronal sources depends on the model of artifact sources, and vice versa [2]. This poses a problem, as an exact model for one cannot be produced in the presence of the other. The PCA model method [2] is included in the SPM package [38].

**Gap filling**

This is an interpolation method discarding the signal with artifact, treating it as totally corrupted and assigning it to gap, and then filling the gap with an appropriate data driven model (see Fig. 1). The edges of the gap depend on the type of artifact, and in [33] the segment [-10, 30] ms around the TMS onset was assigned to gap. The interpolation model proposed in [33] is a local state space model stemming from the nonlinear analysis of time series under the perspective of dynamical systems theory. The hypothesis is that there is a dynamical system underlying the signal from the electrode, and the missing signal can be reconstructed by a model of this dynamical system extrapolating it forwards and backwards in time from the beginning and end of the gap, respectively.

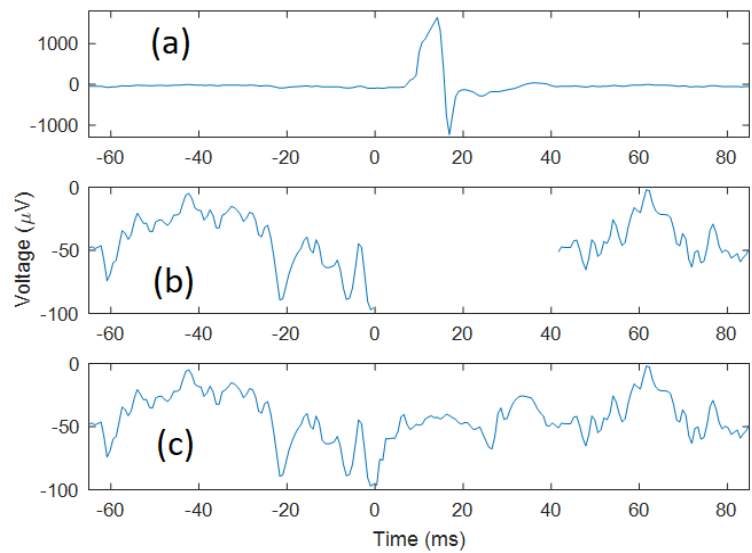

*Figure 1: An EEG segment demonstrating the gap filling method having the TMS onset at time zero: (a) signal with TMS artifact, (b) the corrupted part of the signal of (a) in the time interval [-10, 30] ms is treated as a gap, (c) the missing part is filled by weighted forward and backward prediction using the time series on the left and right of the gap, respectively. Note that the signal outside the gap is the same but in (a) it is at a much larger scale. The scale of voltage in (a) is broader to display the whole artifact being 10 times larger in amplitude than the neuronal activity. The depicted unfiltered EEG signal is plotted with negativity as an downward deflection. This polarity convention is followed in all subsequent figures.*

The model requires the reconstruction of the state space from the measurements of the variable $x_t$, where $x$ is the brain potential measured at an electrode and $t$ is the time index (multiple of the sampling time). The point at time $t$ in the reconstructed state space of dimension $m$ is $\boldsymbol{x}_t = [x_t \ x_{t-\tau} \ ... \ x_{t-(m-1)\tau}]$, where τ is the time delay parameter. In [33], an embedding dimension of $m = 50$ and delay τ = 1 were selected, and the number of neighbors *k* defining the neighborhood around $\boldsymbol{x}_t$ for which the model is valid was set to *k*=2. The prediction of the next observation $x_{t+1}$ is given by the average of the one-time ahead observations of the *k*=2 nearest neighbors of $\boldsymbol{x}_t$. The prediction is iterated for the target point $\boldsymbol{x}_{t+1}$, formed in the same way and given the predicted observation $x_{t+1}$, and until the whole missing segment is estimated, covering the interval [-10,30] ms with reference to the stimulation time point. A very small *k*=2 is used to preserve the local approximation in a very high dimensional space ($m = 50$), and *k*=1 is avoided to preclude predicting reoccurring segments [39]. We let τ = 1 and $m = 50$ to include all details of the fluctuating signal of 50 time units long in the search for similar preceding segments (nearest neighbors). For the prediction of the missing segment the nearest neighbors are sought in the last 210 ms, i.e. the interval [-220, -10] ms (forward prediction). The same missing segment is predicted again by shifting the direction of time and using the interval [30, 280] ms to search for nearest neighbors (backward prediction). The forward and backward predictions for each predicted observation in [-10,30] ms are weighted and the weights change linearly going from 1 to 0 for the forward prediction and from 0 to 1 for the backward prediction as the prediction time goes from -10 ms to 30 ms, so that the sum of the weights always gives 1. There may be considerable changes in the signal before and after the gap, and by this weighting scheme the algorithm adjusts the gap filling to the closest state, before and after the gap [33].

## Data

The TMS-EEG data were recorded at the Lab of Clinical Neurophysiology, the Medical School of Aristotle University of Thessaloniki, from a 37-year-old female patient with Juvenile Absence Epilepsy, a subtype of Genetic Generalized Epilepsy. Recordings were performed in an electrically-shielded room according to TMS-EEG methodological guidelines [4]. The EEG recordings of 60 channels in 10-10 montage were recorded with a TMS compatible EEG system (eXimia, Nexstim Ltd) at a sampling rate of 1450 Hz and bandpass filtered between 0.1 and 500 Hz. Brain stimulation was performed with a Magstim Rapid2 magnetic stimulator (The Magstim Company Ltd) with a circular coil centered over the vertex, which employs a sample-and-hold circuitry to eliminate the TMS artifact [20]. Still, the TMS administration gives rise to high amplitude sharp fluctuations up to 20 ms post-TMS, masking the brain activity in the EEG. The data were processed in Matlab (Mathworks, Matlab R2012a and R2016a) using the FastICA and SPM8 packages.

The sampling frequency and the cutoff frequency of the pre-filter affect the shape of the TMS artifact. The TMS artifact fluctuations have sharp slopes and thus contain considerable power in higher frequencies (>100 Hz). However, if a slow sampling frequency, or equivalently a low cutoff frequency is used, the artifact is reduced in power but also spread over time. To avoid this, a sampling frequency at least 5 times the maximum frequency in the artifact spectrum and a cutoff frequency not below 1000 Hz are recommended [17]. On the other hand, the stretch in time could be corrected since a big part of the artifact power will be gone by the filtering that caused the stretch. The TMS artifact correction algorithms were applied to the raw EEG without any filtering. Filtering before applying the correction algorithms was also tested. In this case, a bandpass filter at 1-100 Hz was used, since the study is not limited to a particular brain rhythm.

### *Artifact estimation and pilot signals*

To estimate the TMS artifact, the average is taken over nine TMS epochs spanning the time interval [-10, 30] ms with reference to the stimulation onset. All TMS epochs are taken from the first TMS pulse in a train of TMS pulses, repeated throughout a recording at rest. The train comprises a block of two TMS pulses at 4 Hz frequency followed after 1 s by a block of 5 TMS at the same frequency, as shown in Fig. 2. Note that the subsequent TMS epochs in the same TMS block may contain effects of the previous TMS pulse [40], and therefore only the first TMS in the block is considered. The artifact differs across channels, so the addition of TMS artifact is channel specific.

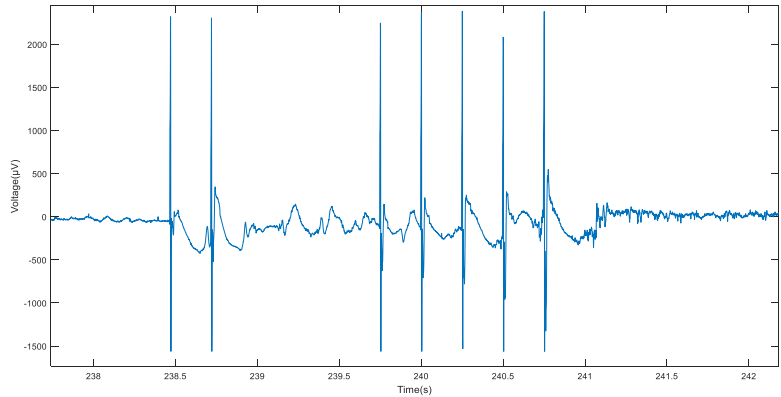

*Figure 2: An EEG segment containing a pattern of 2-block TMS followed by a 5-block TMS.*

The immediate neuronal reaction to the stimulation (TMS-evoked activity) is likely to be similar across blocks and much smaller in amplitude than the artifact. Any other brain activity (spontaneous EEG activity) is expressed as fluctuations around the baseline in the EEG signal and they are expected to cancel out across blocks and not add any systematic pattern in the interval of the average TMS artifact.

The artifact in our measurements appears within the first 30 ms after the pulse, but after estimating it with the aforementioned method we find that at its margins the signal amplitude is still about 3 times larger than that of the EEG at rest. To avoid discontinuities after the artifact is added to the pilot signals, which will introduce high frequencies in the signal and affect filtering, the calculated artifact is multiplied by a steep Tukey window to fix its marginal values at - 10 and 30 ms to zero.

The average TMS artifact spanning a time interval of 40 ms, regarding the period of interest [-10, 30] ms, is added to EEG signals of three types: a) at rest, b) in the beginning of an epileptiform discharge (onset ED), and c) within the ED (mid-ED). The selection of the types b) and c) is motivated by settings that have occurred in studies of our group, namely induction of ED by TMS for b) [14,41], and abortion of ED by TMS for c) [33,42]. The length of each EEG signal on which the TMS artifact is added is 460 ms (669 observations) to account for practical constraints, where the TMS onset is at 210 ms to allow for baseline activity estimation prior to TMS administration and the 250 ms after TMS is in accordance with repeated TMS at a frequency of 4 Hz.

## Results

In this section we assess the quality of TMS artifact correction with reference to the true EEG for each of the three studied methods.

### *FastICA*

The rationale of FastICA method is to identify and remove ICs corresponding to the TMS artifact. Initially, the method was applied to all but Cz channels and without PCA dimension reduction and gave 59 ICs. An example of five randomly chosen EEG signals and ICs is shown in Fig. 3a and d.

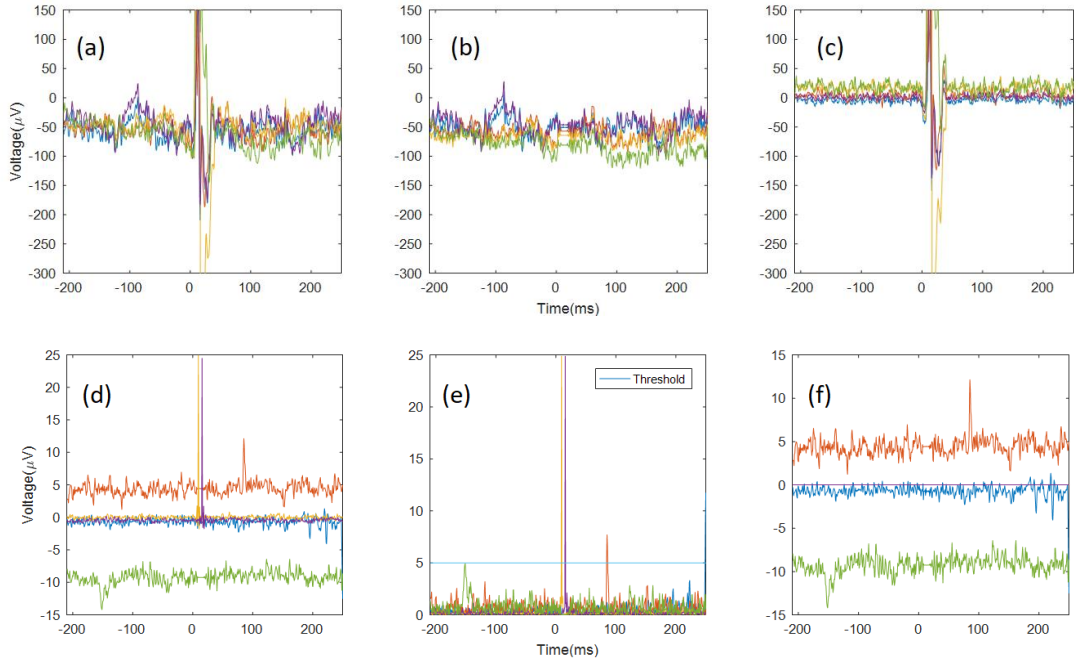

*Figure 3: The TMS artifact correction with FastICA on 59 EEG channels: (a) five randomly selected original EEG signals, (b) reconstructed EEG signals after removing ICs containing strong TMS artifact, (c) the difference of EEG signals before and after TMS artifact correction, (d) five randomly selected ICs, (e) absolute value of ICs with the horizontal line denoting the threshold, (f) three of the 31 ICs having amplitudes less than the threshold in (e) (note the different voltage scale).*

It appears that many ICs signals exhibit peaks of varying amplitude at the time of TMS and thus it is hard to identify ICs containing the TMS artifact and others containing only brain activity. We set a threshold for the amplitude to identify the ICs having stronger TMS artifact elements, as shown in Fig. 3e. This threshold was found by simple visual inspection of the ICs and was set to 5 μV for TMS during resting state and 6.5 μV otherwise, since these values always separated ICs with a strong component during the TMS pulse. The 31 remaining ICs having amplitude smaller than the threshold, three of which are shown in Fig. 3f, seem to lack any activity at the time of TMS. The 59 EEG signals reconstructed from the 31 ICs of low amplitude, shown in Fig. 3b, also do not exhibit any neuronal activity at the time window of the TMS artifact and display merely a straight line. On the other hand, the discarded ICs of large amplitude do not go to zero at times outside the TMS window, indicating that they carry some information about neuronal activity as well. Apparently, the artifact cannot be isolated from the rest of the neuronal activity in any subset of ICs. As a result, the reconstructed EEG signals exhibit significantly deformed neuronal activity outside the TMS window, as can be seen from the large difference in EEG before and after the artifact correction in times outside the TMS window, shown in Fig. 3c.

We apply FastICA employing PCA dimension reduction using the 'symmetric' approach and retain 15 ICs, which is the maximum that can be obtained for waveforms of 669 observations according to the $3N^2$ rule. We apply the same thresholding resulting to only two ICs of low amplitude and thus the 59 reconstructed EEG signals from these two ICs are heavily correlated and look very similar. The very large amplitude of the TMS artifact results in almost all ICs being required to describe the variability of the TMS artifact, leaving only few ICs (two here) to be more representative of the neuronal activity. The opposite would be desired, i.e. few ICs to represent the TMS artifact and be further discarded, but the algorithm cannot attain this.

In an attempt to assess the effectiveness of the algorithm if the TMS artifact would not have such large amplitude, we selected for TMS artifact correction a subset of 28 EEG channels having the smallest amplitude of TMS artifact, excluding some irrelevant channels in perimeter areas, as well as nearby channels with almost identical artifact, so as to include channels from all scalp regions. We applied PCA dimension reduction to 15 ICs using the 'symmetric' approach and after thresholding we obtained three ICs. The 'deflation' approach gave four ICs. In both cases the 28 reconstructed EEG signals had distinctly larger amplitudes at the time window of the TMS artifact, indicating that the TMS artifact was not successfully corrected, and moreover the TMS artifact was not isolated and a great deal of neuronal activity outside of the TMS window was lost.

In an effort to obtain more ICs without TMS artifact components, EEG channels were further reduced to 18 channels relaxing the criterion of having channels from all scalp regions. Again, using PCA dimensional reduction to 15 ICs, the symmetric approach and thresholding, we obtained four ICs and the reconstructed EEG signals were distinctly different at times outside the TMS artifact, indicating that the method is unable to separate the neuronal activity from the TMS artifact.

In a last attempt to find positive results with this method, we first filtered the EEG signals with a low pass filter at 100 Hz. The benefit of this is that TMS artifact amplitude is reduced as it contains frequencies higher than 100 Hz, while neuronal activity is not altered as it is mostly found in frequencies lower than 100 Hz. Also, the filtering of the background noise at high frequencies increases the efficiency of the PCA dimension reduction. However, the results did not improve. FastICA did not converge when 18 channels were selected and the reduced dimension was set to 15, and therefore no dimension reduction was applied. When the 'deflation' approach was used, the reconstructed EEG signals again differed significantly from the original ones outside the period of the TMS artifact.

So far, we have seen results from resting EEG. During an epileptic seizure, EEG is much larger in amplitude, which makes the difference between neuronal and artifact signals less prevalent. As seen in Fig. 4a for the case of 18 EEG signals, large peaks outside the [0, 40] ms window, corresponding to the epileptic activity, are reproduced satisfactorily by the FastICA method (PCA dimension reduction set to 15 ICs and 'deflation' approach was used, as the 'symmetric' approach did not converge). These large epileptic fluctuations are represented well by some ICs. However, the error amplitude remains at the order of neuronal activity at rest. Moreover, as shown in Fig. 4b, the correction of the TMS artifact leads to a high frequency oscillation of smaller amplitude in the artifact window because again the discarded ICs assigned to TMS artifact carry information on neuronal activity during the ED. These results are similar to the ones obtained with the same procedure (the 'symmetric' approach is shown here since it converged) on the third EEG signal type having TMS in the beginning of ED, as can be seen in Fig. 4c and d.

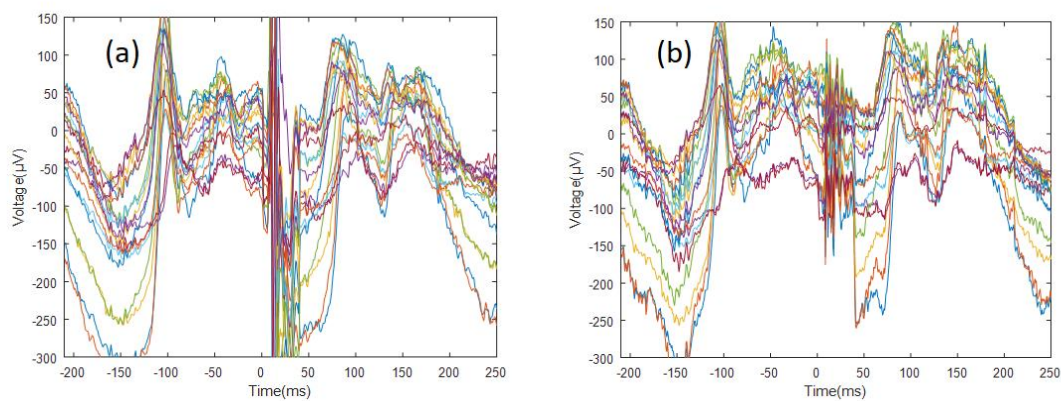

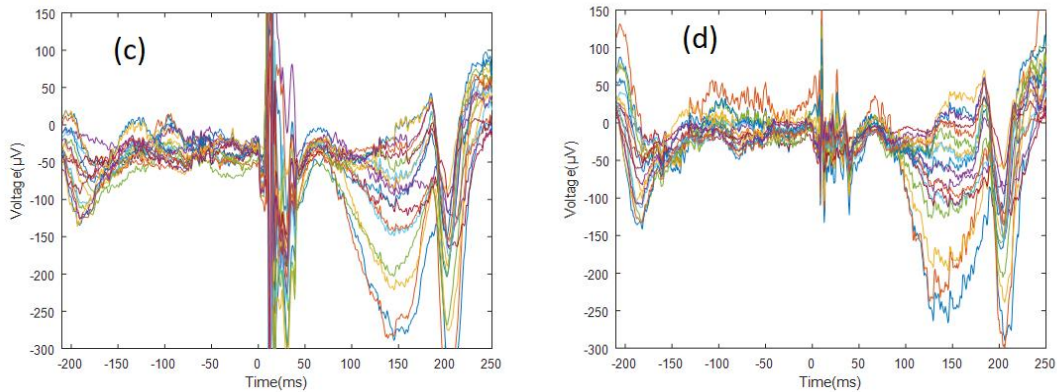

*Figure 4: FastICA results for 18 channels and dimension reduction to 15 for the TMS administration during ED with the TMS artifact in (a) and with the TMS artifact corrected in (b), and for the TMS administration inducing ED with the TMS artifact in (c) and with the TMS artifact corrected in (d).*

### PCA model

The application of FastICA,First showed that the TMS artifact has large amplitude varying across channels, so that the artifact cannot be captured by a small subset of components of the ICA applied to the set of EEG signals. The same is expected to apply for PCA. The PCA model method uses PCA expecting that the artifact will be expressed only in the first principal components. This does not seem to be the case, as confirmed by our results on the set of 55 channels. For the resting state EEG signals, the method only reduces the TMS artifact amplitude to a lesser or larger degree (Fig. 5a and b, respectively). The reduction of the TMS artifact comes also at the cost of distortion of the signal outside the TMS artifact.

The distortion of the signal outside the TMS artifact was larger for the other two scenarios of TMS on ED and TMS inducing ED, and the amplitude of the TMS artifact was less reduced compared to the resting state (Fig. 5c). This is also attributed to the different intensity and shape of the TMS artifact across the 55 channels.

In an attempt to modify the setting of the TMS artifact in favor of the PCA model we selected 18 of the 55 channels having the lowest TMS artifact amplitude and similar shape. The results for resting state (Fig. 5d) were slightly better as there was a better suppression of the artifact, with only small increases in a few channels, but the same problems remained.

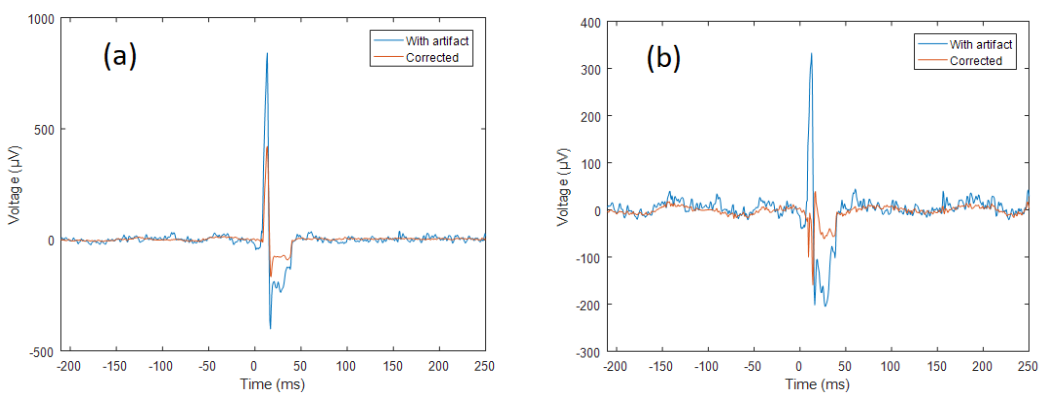

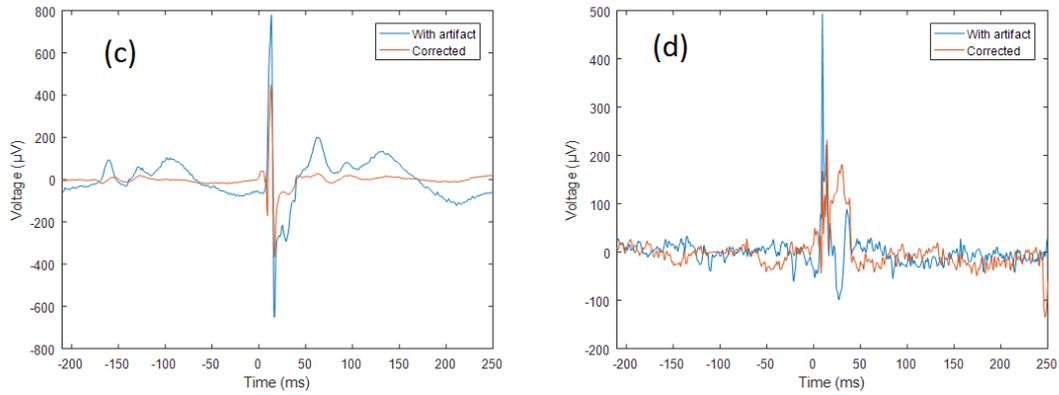

*Figure 5: The EEG signal with TMS artifact and the corrected signal from the PCA model method applied to 55 EEG channels: (a) CP3 signal at resting state, (b) P4 signal at resting state, (c) C1 signal with TMS during ED. In (d) the raw and corrected FP2 signal at resting state is shown where the PCA model method is applied to the 18 channels having low artifact amplitude.*

### *Gap filling:*

In contrast to the FastICA and PCA model artifact correction, the gap filling method gave satisfactory results on resting state EEG signals. In general, the oscillations of the neuronal activity masked by the TMS artifact were closely approximated, in cases succeeding also to be in phase (Fig. 6a) and in other cases with a time lag (Fig. 6c). Less frequently, it failed to capture the original oscillations at a lesser or larger extent (Fig. 6e). To show the correction more clearly in the three representative examples in Fig. 6, the same signals are shown after low pass filtering at 100 Hz in Fig. 6b, c and d, respectively.

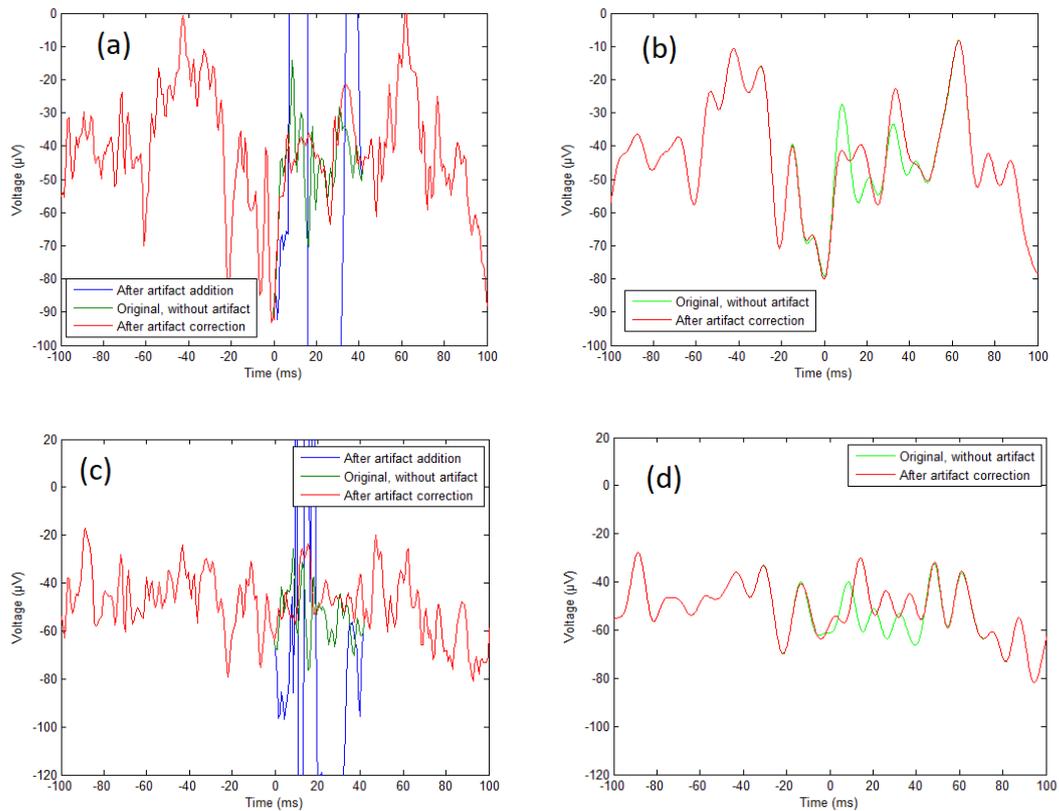

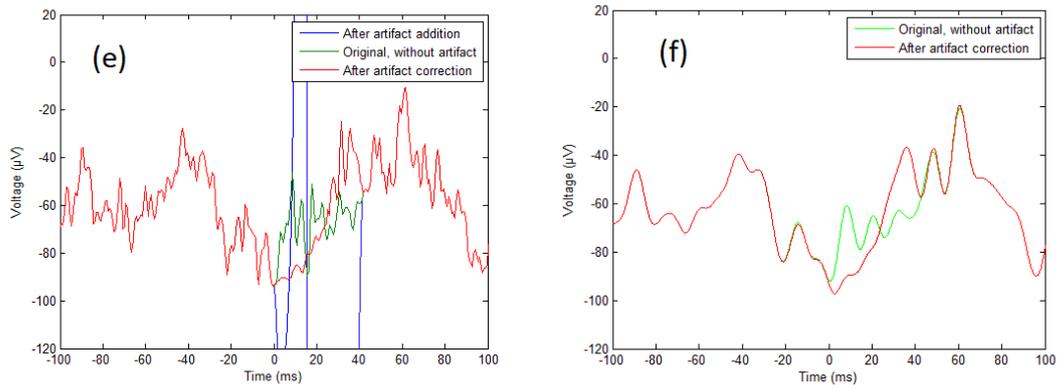

*Figure 6: Three examples of TMS artifact correction with gap filling demonstrating different types of approximation of the true EEG signal. Each panel shows the true EEG signal, the EEG signal with the superposition of the TMS artifact, and the corrected signal, as shown in the legend in (a), (c) and (e): (a) EEG signal F5, (c) EEG signal T3 and (e) EEG signal C4. In (b), (d) and (f) the true and corrected EEG signals are shown as in (a), (c) and (e), respectively, but after low pass filtering at 100 Hz.*

The results from artifact correction during epileptiform discharge (ED) showed that unlike the case of resting EEG the corrected signals may deviate in shape substantially from the original ones. While in some cases the corrected signal has similar shape to the original one (Fig. 7a), in many other cases, peaks appear in the corrected signal not existing in the original one (Fig. 7b). This is to be expected, as seizures last for a few seconds and the large amplitude oscillations have period of about 1/4 of a second, being half the time window of the training set for the gap filling model. Even when the deviation appears to be relatively small (Fig. 7a), it is still large compared to the rest EEG case. Thus, the artifact correction cannot approximate the original EEG signal but nevertheless the produced signal by the gap filling method contains oscillations that could be considered as part of the ED activity.

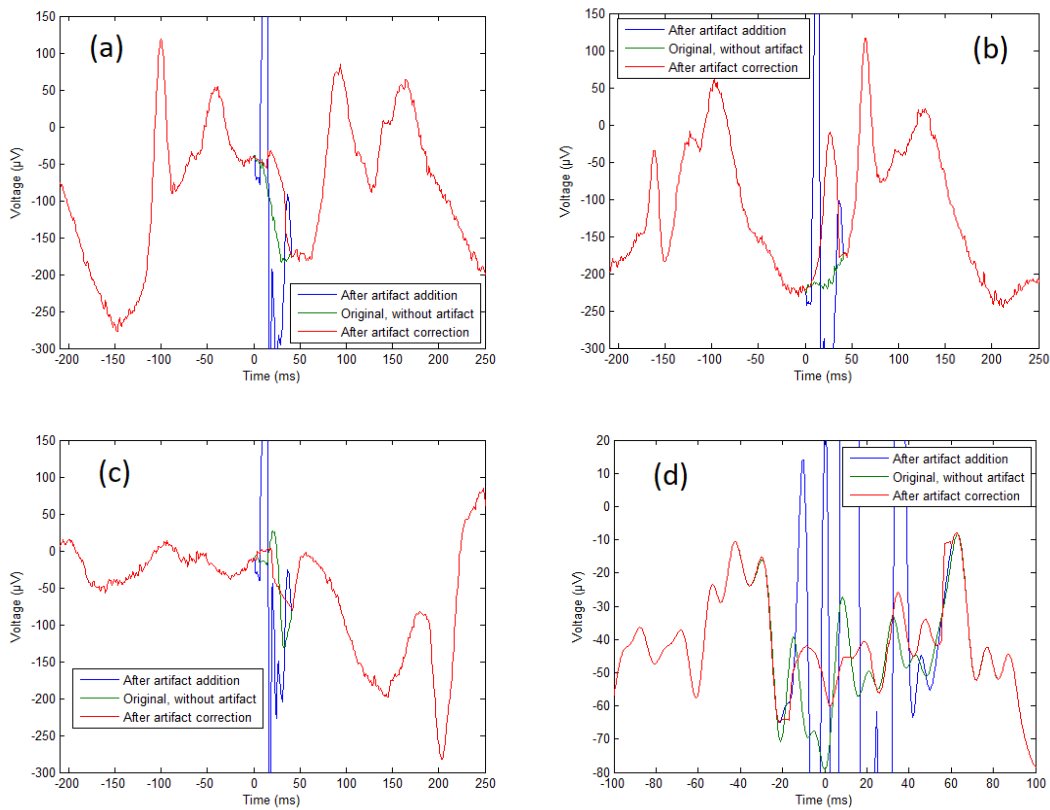

*Figure 7: Two examples of TMS artifact correction with gap filling when TMS is administrated during an ED in (a) and (b) and one example when TMS is administered at rest and induces ED in (c). In (a), (b) and (c) the EEG signal is from F5. In (d), the signal at resting state shown in Fig.6a is first low pass filtered at 100 Hz, resulting in longer duration of the artifact, and then corrected.*

The onset ED signal (TMS-induced ED) has similar features with the mid-ED signal (slow waves) at the part after the magnetic stimulus and therefore it is likewise difficult to retrieve the masked signal by the TMS artifact. However, as the first part before the TMS is at rest, the deviation from the true EEG signal was smaller than for the mid-ED signal, as shown in a representative example in Fig. 7c.

Filtering the TMS corrupted EEG signal prior to applying the gap filling method could possibly improve the predictability of the method. However, filtering increases the duration of the artifact. As shown in Fig. 7d, the TMS artifact after filtering extends in both edges and the duration goes from 40 ms before filtering to about 70 ms after filtering. Thus, the gap to be filled by the forward-backward prediction method is longer and the overall approximation of the true signal is worse, as shown in the example of the filtered signal in Fig. 7d, to be compared to the raw signal in Fig. 6a and b. The increase in TMS artifact duration with filtering depends on the artifact amplitude. For the 100 Hz cutoff filtering the variance of the artifact is reduced by a factor of about 1.4. Using a larger cutoff results in smaller variance reduction and artifact duration increase, e.g. for 200 Hz cutoff the duration is at about 60 ms and the correction is still worse than that of the raw signal.

Exact predictions of such a complex signal cannot be anticipated by any interpolation method and the question of interest would rather be whether the method can fill the gap with neuronal-like activity relevant to the brain state in this time window. For this, there is no direct way to assess the quality of correction of the TMS artifact with the gap filling method even when the true masked signal is known. Indeed, the method is not meant to provide an accurate prediction of the true signal, so that one can quantify the prediction error. However, a measure of prediction error can still indicate whether the shape of the signal (amplitude, period) is well preserved. We compute the root mean square error (RMSE), i.e. the square root of the mean of the squares of the differences between predicted and true signal values in the window of correction for every channel and epoch. The RMSE is not always an indicative measure of the preservation of the neuronal-like activity in the corrected signal, e.g. the correction may be the same to the original signal but with a time lag and then RMSE is large (see Fig. 6d). To this respect, the normalized RMSE (NRMSE), defined by dividing RMSE with the SD of the true signal, can be a better indicator as it quantifies the deviation from the mean prediction (for the mean prediction NRMSE=1). Thus, if the shape of neuronal activity is preserved we expect to have NRMSE close or only slightly higher than one and if other characteristics of larger amplitude appear in the corrected signal, such as peaks, NRMSE would be much larger than one. The average RMSE and NRMSE for every channel and epoch for the three states and for raw and filtered signal are shown in Table 1.

|  | Resting EEG | Onset ED | Mid-ED |
|---|---|---|---|
| RMSE ($\mu$V) - no filter | 13.95 | 20.23 | 44.29 |
| NRMSE – no filter | 1.353 | 1.140 | 2.651 |
| RMSE ($\mu$V) - Lowpass at 100 Hz | 11.00 | 19.73 | 43.86 |
| NRMSE - Lowpass at 100 Hz | 1.479 | 1.192 | 2.711 |

*Table 1 – Average RMSE and NRMSE for the gap filling correction of EEG signals of the three different types, with and without a filter after correction.*

The filtering gives significant improvement of RMSE in resting EEG due to the reduction of noise. There is no significant improvement in onset ED and mid-ED signals, since the same level of noise is small compared to the signal amplitude. The correction for the resting EEG is far better than for mid-ED and better than for the onset ED EEG and this is apparently due to the smaller amplitude of the signal in resting EEG. For the same reason the correction for the onset ED EEG is better than for the mid-ED EEG (the first part of onset ED EEG has much smaller amplitude than for the mid-ED EEG). The NRMSE values are close to one for the resting EEG and onset ED EEG indicating that the corrected signal has similar fluctuations to the original one. However, for the mid-ED EEG, the often-observed peaks in the corrected signal that do not exist in the true EEG signal results in much larger NRMSE. There is a slight increase of NRMSE with filtering for all three signal types because the corrected signals tend to be smoother and thus closer to the mean prediction.

For this method we explore two free parameters, the length $w$ of the time window before and after the gap and the embedding dimension $m$. Regarding $m$, the reason for initially selecting $m=50$ is to have long enough information from the past in the regressor vectors for the prediction in longer horizon to be more accurate. Moreover, a large $m$ allows for better representation and estimation of longer duration signals, such as the epileptic seizure cycle of about ¼ s. However, the span to the past determined by $m$ is in dependence to the length of the time series the training data are formed from, here defined by $w$. As the RMSE and NRMSE for both filtered and non-filtered mid-ED EEG signals show in Table 2, for the relatively small $w=460$ used in the original study in [33] and here as well, the choice $m=25$ gives fluctuating signals for the gap closer to the true EEG signal. On the other hand, increasing $m$ to 100 does not change the RMSE and NRMSE figures, but we have experienced more often false peaks in the gap filled signals. It is noted that the choice of the time window of duration 250 ms after the gap was imposed by the use of block TMS with a frequency of 4 Hz (250 ms). We repeated calculations with the gap filling method on mid-ED EEG for larger $w=2400$ ms. The RMSE and NRMSE for $m=50$ got smaller than for the small $w$ and for larger $m$ (up to 300) the results do not change (see Table 2). Thus, if the stimulating setting allows the use of a larger data window before and after the TMS, e.g. single-pulse TMS or TMS blocks of lower frequency, the fluctuating signals filling the gaps can be closer to the true EEG mid-ED signals in shape. Repeating the same experiment on onset ED EEG the fitting was somewhat worse. For the window length $w$, we also tried other values between 460 ms and 2400 ms and for the same $m$ we did not observed any improvement in the quality of the TMS artifact correction. For example, when we doubled the data window (from left and right of the TMS artifact) for the resting EEG signal the RMSE was about the same as for the standard window length.

| Type | Parameters | | No filter | | Filter at 100 Hz | |
|---|---|---|---|---|---|---|
| | $m$ | $w$ | RMSE | NRMSE | RMSE | NRMSE |
| Mid-ED | 25 | 460 | 34.46 | 2.10 | 34.17 | 2.15 |
| Mid-ED | 50 | 460 | 44.29 | 2.65 | 43.86 | 2.71 |
| Mid-ED | 100 | 460 | 44.37 | 2.90 | 43.88 | 3.00 |
| Mid-ED | 50 | 2400 | 34.81 | 2.80 | 34.74 | 2.73 |
| Mid-ED | 300 | 2400 | 35.92 | 2.86 | 35.52 | 2.69 |
| Onset ED | 50 | 2400 | 41.76 | 1.49 | 41.53 | 1.53 |

*Table 2 – Average RMSE and NRMSE with and without filtering after signal correction for varying values of the free parameters in the gap filling method, the window length w and the embedding dimension m.*

## Discussion

In the present study, three methods for correction of the TMS artifact on EEG signals (i.e. FastICA, the PCA model and the gap filling method) were tested and compared on the same set of high-density EEG signals. The EEG signals were constructed by superimposing a representative TMS artifact signal on the raw EEG signal. The TMS artifact signal was extracted as the average of an ensemble of aligned EEG segments at resting state containing a single TMS pulse. The tested EEG signals had the TMS artifact at resting state, during epileptiform discharge (ED) and at the beginning of the ED (as if TMS were inducing the ED).

Both FastICA and the PCA model use a transform on the set of signals, ICA and PCA, respectively. It turned out that the TMS artifact was expressed in many components of ICA and PCA in a non-systematic manner, so that it could not be isolated by a small subset of components. Therefore, both FastICA and PCA model methods could not correct the TMS artifact and the corrected signal still contained the artifact (in smaller amplitude) and in addition the rest of the signal outside the TMS artifact window was also distorted. For FastICA, many ICs were removed to reduce the TMS artifact, but these ICs contained also neuronal activity, so that either the TMS artifact was not substantially reduced or the corrected EEG signal outside the artifact deviated substantially from the true signal. The PCA model method had similar problems. Specifically, the effectiveness of the algorithm is based on the assumption that the artifact is fully recovered by the first few PCs [2], which was not the case with our data of TMS artifact. Like FastICA, the PCA model method reduced the amplitude of the TMS artifact, while simultaneously under-representing neuronal activity outside the artifact window, e.g. the peaks occurring during ED were smoothed out.

The third method, the gap filling method, is applied to each channel independently and unlike the other two methods it does not attempt to correct the TMS artifact but rather it assigns it to a gap and attempts to predict the signal in the gap. The prediction is based on the segments on the left and right of the gap combining a forward and backward prediction with a local state space method. This method by construction does not distort the signal outside the TMS artifact and does not preserve any feature of the TMS. Indeed, for our tested EEG signals corrupted by TMS artifact the gap filling method gave quite satisfactory results and the predicted signal during the period of the TMS artifact bore some similarity to the true EEG signal. Thus the gap filling algorithm offers a far better alternative in relation to linear interpolation, which distorts the spectral content of the signal and replaces the gap with a signal that is not EEG-like.

The results on the tested EEG signals should be treated with caution. By simply superimposing the TMS artifact to the EEG signals, the possible causal relationship between the artifact and the induced neuronal activity is not preserved. However, excluding this property makes the data setting simpler. If a method is not capable of isolating the TMS artifact under these favorable conditions, then it will be unable to isolate the TMS artifact when trying to correct actual data, that is, when a causal coupling between the TMS artifact and neuronal activity does exist. For FastICA that assumes statistical independence between artifact and neuronal activity, our setting should be particularly favorable, but still FastICA did not succeed to correct the TMS artifact.

Also, the fact that the gap filling algorithm performs well here is not an indication that it will also perform well when there is a strong causal relationship between artifact and induced neuronal activity. The hypothesis when using this algorithm is that there is no excess information in the window of the TMS artifact to be corrected, so this part of the signal can simply be replaced by the predicted signal using information from the rest of the signal. So, if the neuronal activity after the period of the TMS artifact is changed as a result of the causal effect of TMS, stationarity for the whole signal (including the part replaced with a gap) is not established and the method may fail.

Estimating the efficiency of an artifact correction algorithm is a process complicated by the issue of finding the ground truth. If the ground truth of neuronal activity is not known, the output of the method cannot be evaluated. The EEG signals employed in this work, containing a superimposed TMS artifact to EEG signals at different states, represent an attempt to address this issue. An alternative approach would be to reside in qualitative measures, using time-frequency diagrams [34]. The gap filling algorithm cannot promise to find the discarded signal, but its rationale is to replace the gap with a signal with statistics similar to that of EEG. Whereas ICA and PCA tend to eliminate oscillations, the gap filling algorithm preserves them, without necessarily preserving their phase. If, however, we were to receive the Fourier transform of the corrected signal, it would closely resemble an EEG signal without the TMS artifact. So, one approach for developing a gap filling correction algorithm would be to minimize the distortion of the spectral content of the EEG signal, which is a topic for further research.